\documentclass[journal]{IEEEtran}
\usepackage{graphicx}
\usepackage{caption}
\usepackage{subcaption}
\usepackage{float}

\begin{document}

\title{Construction and Test of New Precision Drift-Tube Chambers for the ATLAS Muon Spectrometer}

\author{\underline{H.~Kroha$^*$}\thanks{$^*$Corresponding author: kroha@mpp.mpg.de}, O.~Kortner, K.~Schmidt-Sommerfeld, E.~Takasugi
 \\ \textit{Max-Planck-Institut f\"ur Physik, F\"ohringer Ring 6, D-80805 Munich, Germany}}

\maketitle
\pagestyle{empty}
\thispagestyle{empty}

\begin{abstract}	
	ATLAS muon detector upgrades aim for increased acceptance for muon triggering and precision tracking and for improved rate capability of the muon chambers 
	in the high-background regions of the detector with increasing LHC luminosity. The small-diameter Muon Drift Tube (sMDT) chambers have been developed for these purposes.
	With half of the drift-tube diameter of the MDT chambers and otherwise unchanged operating parameters, sMDT chambers share the advantages of the MDTs, but have an order 
	of magnitude higher rate capability and can be installed in detector regions where MDT chambers do not fit in.
	The chamber assembly methods have been optimized for mass production, minimizing construction time and personnel. 
	Sense wire positioning accuracies of 5~$\mu$m have been achieved in serial production for large-size chambers comprising several hundred drift tubes.  
	The construction of new sMDT chambers for installation in the 2016/17 winter shutdown of the LHC and the design of sMDT chambers in combination with new RPC trigger chambers
	for replacement of the inner layer of the barrel muon spectrometer are in progress. 
\end{abstract}

%\linenumbers

\section{sMDT Precision Muon Tracking Chambers}

The Monitored Drift Tube (MDT) chambers of the ATLAS muon spectrometer~\cite{ATLAS} have demonstrated that they provide very precise and robust tracking over large areas. 
Goals of ATLAS muon detector upgrades are to increase the acceptance for triggering ande precise muon momentum measurement  
and to improve the rate capability of the muon chambers in the high-background regions at increasing LHC luminosity.
Small-diameter Muon Drift Tube (sMDT) chambers have been developed for these purposes. With half the drift-tube diameter of the MDT chambers (15 instead of 30 mm) and otherwise 
unchanged operating parameters, including a gas gain of 20000 in Ar:CO$_2$ (93:7) gas mixture at 3 bar, 
sMDT chambers share the advantages of the MDTs, but have an order of magnitude higher rate capability~\cite{sMDT} 
and can be installed in detector regions where MDT chambers do not fit in.

The construction of twelve sMDT chambers to be installed in the feet of the ATLAS detector in the middle layer of the barrel muon spectrometer 
in the LHC winter shutdown 2016/17 is in progress since October 2014. They consist of 356 drift tubes of 1120~mm length in two quadruple layers (multilayers) which are 
separated by an aluminum spacer and support frame. Half of the chambers has already been constructed.
The purpose of this upgrade is to increase the acceptance for the three-point measurement of muon tracks which will result in an substantial improvement of
the muon momentum resolution. Two similar chambers with 624 tubes of 2150 mm length have already been installed in 2014 in the middle layer of the bottom barrel sector of the spectrometer 
for the same purpose. 
For operation at HL-LHC, replacement of the MDT chambers in the barrel inner layer by sMDT chambers with integrated thin-gap RPC chambers is foreseen in order to increase the
muon trigger acceptance and robustness as well as the rate capability of the precision tracking chambers. At the same time the drift distance information from the MDT and sMDT 
chambers will be included in the first-level muon trigger. In the LHC shutdown 2019/20, 16 such sMDT and RPC chamber packages will already be installed at the ends of the barrel 
inner layer for LHC run 3.

\section{Chamber Design and Construction}

\begin{figure}[h!]
	\centering
	\begin{subfigure}[b]{0.45\textwidth}
	\centering
	\includegraphics[angle=-90,width=\textwidth]{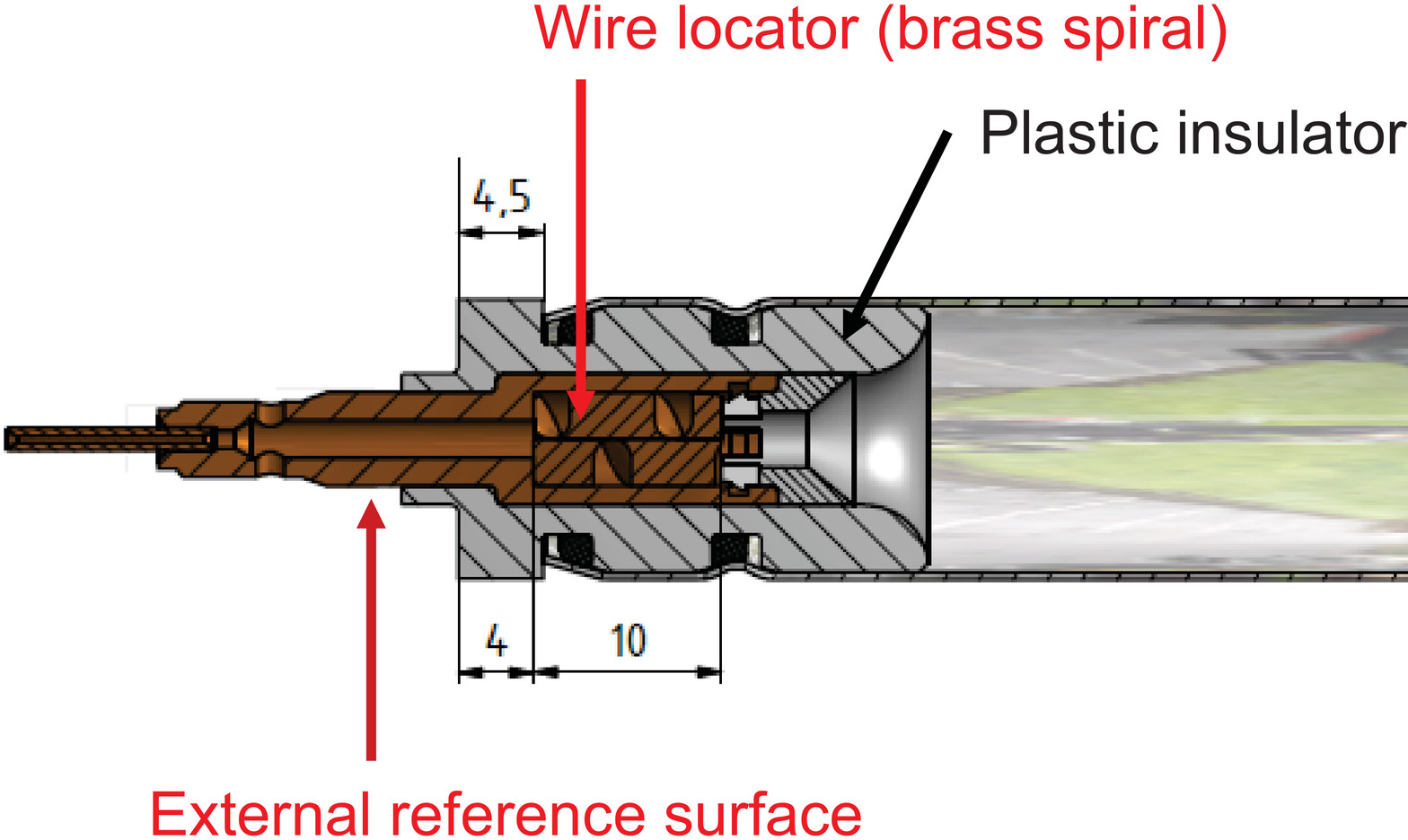}
	\caption{}
	\label{fig::endplug}
	\end{subfigure}
	\qquad
	\begin{subfigure}[b]{0.45\textwidth}
	\centering
	\includegraphics[width=1.14\textwidth]{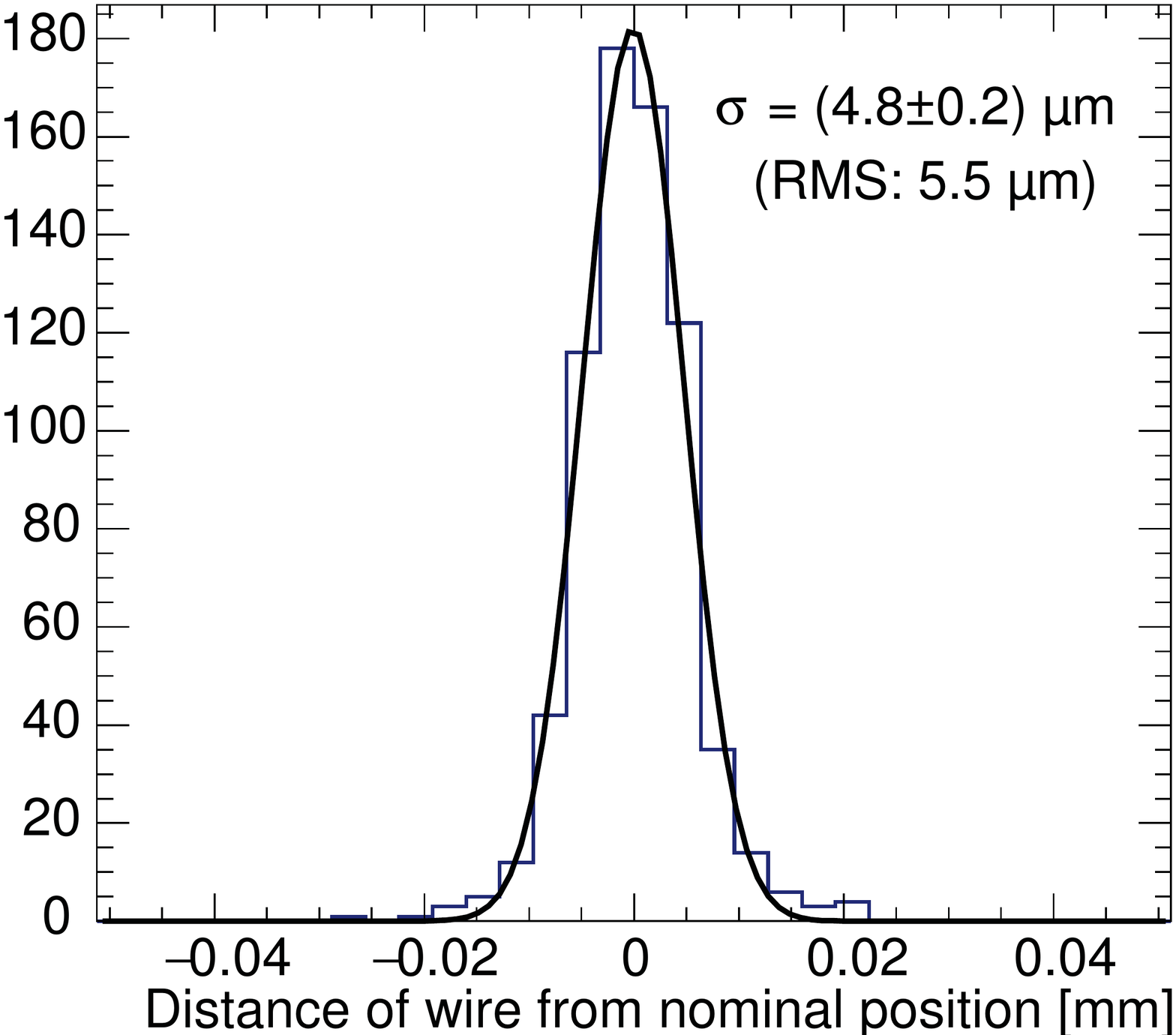}
	\caption{} 	
	\label{fig::residuals}
	\end{subfigure}
\vspace{-3mm}
	\caption{(a) Cross section of an sMDT endplug with internal wire locator and external reference surface for tube positioning and wie position measurement.
	         (b) Residuals of the sense wire positions measured at both ends of an sMDT chamber with 356 tubes with respect to the nominal wire grid. The width of the distribution 
	         includes the measurement accuracy of about 2~$\mu$m.}  
\end{figure}

The chamber design and construction procedures have been optimized for mass production while providing highest mechanical accuracy in the sense wire positioning.
The sMDT drift tubes are standard industrial aluminium tubes with 15~mm diameter and a wall thickness of 0.4~mm, chromatised for cleaning and reliable electrical ground contact.
First, the drift tubes are assembled using a semi-automated wiring machine in a temperature-controlled clean room.
The sense wires are fed through the tubes and the endplugs (see Fig.~\ref{fig::endplug}) without manual contact by means of air flow and fixed in 
copper crimp tubelets after tensioning to $350\pm 15$~g corresponding to a gravitational sag of only $17\pm 1~\mu$m (absolute tolerances).
The wires are positioned at each tube end with few micron precision with respect to a cylindrical external reference surface
on the central brass insert of the endplug which also holds the spiral shaped wire locator on the inside of the tube. 

The drift tubes are sealed with the endplugs via two O-rings each using mechanical swaging of the tube walls. For the injection molded endplug insulators 
and gas connectors for the individual tubes, plastic materials with minimum outgassing have been selected which also are immune against cracking. 
No aging has been observed in irradiation tests of the sMDT drift tubes up to 9 C/cm charge accumulated on the wire.
The 4300 drift tubes of the twelve new chambers have been produced at a typical rate of 100 tubes per day and a failure rate of less than $4\%$
due to intermittent failures of the wire tensioning and crimping tools detected by visual inspection during drift tube assembly.
The critical performance parameters, wire tension, gas leak rate ($<10^{-8}$~bar~l/s) and leakage current ($< 2$~nA/m) at nominal operating voltage (2730~V) 
are measured for every tube before assembly in a chamber. The overall failure rate of these tests was only $0.5\%$. 

The drift tubes are assembled into chambers in a climatised clean room by inserting the endplug reference surfaces into a grid of fitting bores in the assembly jigs at each chamber end
which define the wire positions with an accuracy of better than 5 micron and glueing them together and to the spacer and support frame 
using an automated glue dispenser. 
A complete chamber can be assembled within one working day. The precise mounting of the platforms for the optical alignment sensors with respect to the wires requires an additional day.   
Before the installation of the parallel gas distribution system (which fulfills the stringent leak rate requirement of less than 0.2 mbar/h) and of the specially designed electronics boards with 6 x 4 readout channels matching the transverse cross section 
of the tube multilayers, the positions of the individual sense wires are measured at the two chamber ends via the endplug reference surfaces with an automated coordinate measuring machine
with a precision of about 2~$\mu$m (rms).
At the same time, the relative angle between the wire grids at the two chamber ends (chamber torsion) and the positions and orientations of the optical alignment sensor mounting platforms 
are measured. Sense wire positioning accuracies of better than 10~$\mu$m (rms) are routinely achieved, surpassing the requirement for the ATLAS MDT chambers of 20~$\mu$m (rms) and allowing for 
a spatial resolution of a chamber with 2 x 4 tube layers of better than 35~$\mu$m. In the best cases, the ultimate wire positioning accuracy of 5~$\mu$m (rms) as given by the precision of the 
assembly jigs has been reached (see Fig.~\ref{fig::residuals}). After the measurement, the wire positions are known with 2~$\mu$m accuracy.

\end{document}